# NewSQL: Towards Next-Generation Scalable RDBMS for Online Transaction Processing (OLTP) for Big Data Management

A B M Moniruzzaman
*Department of Computer Science and Engineering,
Daffodil International University*
abm.mzkhan@gmail.com

*Abstract*

*One of the key advances in resolving the "big-data" problem has been the emergence of an alternative database technology. Today, classic RDBMS are complemented by a rich set of alternative Data Management Systems (DMS) specially designed to handle the volume, variety, velocity and variability of Big Data collections; these DMS include NoSQL, NewSQL and Search-based systems. NewSQL is a class of modern relational database management systems (RDBMS) that provide the same scalable performance of NoSQL systems for online transaction processing (OLTP) read-write workloads while still maintaining the ACID guarantees of a traditional database system. This paper discusses about NewSQL data management system; and compares with NoSQL and with traditional database system. This paper covers architecture, characteristics, classification of NewSQL databases for online transaction processing (OLTP) for Big data management. It also provides the list of popular NoSQL as well as NewSQL databases in separate categorized tables. This paper compares SQL based RDBMS, NoSQL and NewSQL databases with set of metrics; as well as, addressed some research issues of NoSQL and NewSQL.*

## 1 Introduction

In this paper is presented as follows: section 1.1, explores background of NoSQL and NewSQL databases systems. Section 2, Discuss NoSQL systems and some bottlenecks for large-scale data management systems as well as, discuss what inspired the NewSQL movement. In the section 3, NewSQL is described. Section 4, discusses the NewSQL Architecture. Section 5, explores the characteristics of NoSQL Databases. Section 6, show the classification of NoSQL Databases into broad three catagories: New Architecture databases; New MySQL storage engines; and Transparent clustering/sharding. Section 7, compares of SQL based RDBMS, NoSQL and NewSQL with matrix. And finally section 8, shows conclusion of this paper.

### 1.1 Background

"Big Data" is typically considered to be a collection of huge data in very high volume, variety and velocity in nature that can not be effectively or affordably managed with conventional data management tools: e.g., classic relational database management systems (RDBMS) or conventional search engines [1]. One of the key advances in resolving the "big-data" problem has been the emergence of an alternative database technology. This would be broad class of DBMS that differ significantly from the classic RDBMS model. These data stores may not require fixed table schemas, usually avoid join operations and typically scale horizontally; where data is stored in a distributed way, but accessed and analysed from applications.

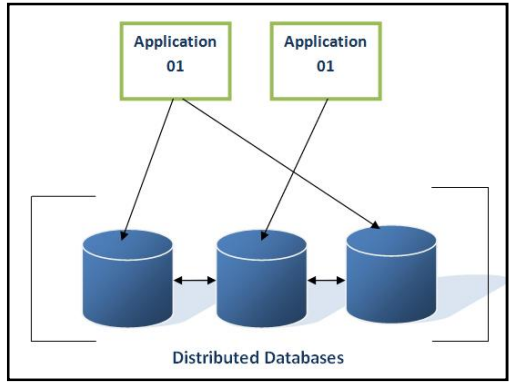
Figure 1.1: Distributed Databases

In the online transaction processing (OLTP), workloads are characterized in small, interactive transactions that generally require high speed response times [2]. High concurrency is also requirements for OLTP systems. The continuing trend of data growth and massive traffic in OLTP systems and, to satisfy a very high availability rate, high responsiveness; a new generation of solutions is required to cater to them. Therefore, DBMSs have to be re-architected from scratch to meet this demand.

## 2 NoSQL

NoSQL stands for "not only SQL" [3]. In broader sense, it includes all non-relational DBMS (which may or may not use a querying language). As opposed to transactions in RDBMS conforming to ACID (Atomicity, Consistency, Isolation, Durability), NoSQL DBMS follow the CAP (Consistency, Availability, Partition tolerance) theorem and thus their transactions conform to the BASE principle [3]. DBMS based on CAP, instead of having their transactions conform to ACID, conform to BASE (Basically, Available, Soft State, Eventually Consistent) properties.

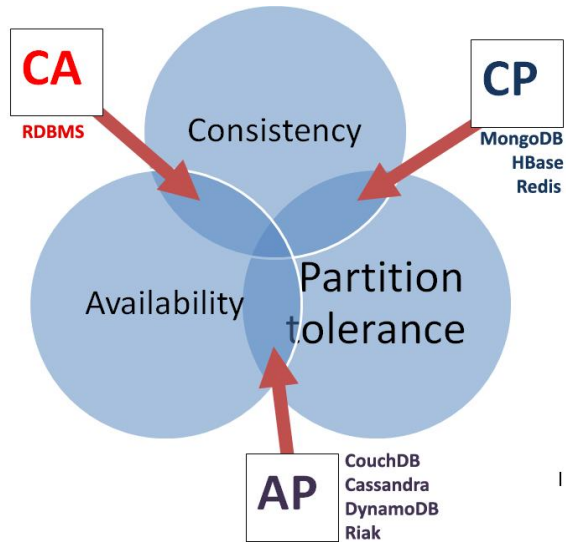

Figure 2.1: CAP theorem with supported NoSQL Databases

NoSQL systems are distributed, non-relational databases designed for large-scale data storage and for massively-parallel data processing across a large number of commodity servers [4]. Types of NoSQL databases include – 1) Key-Value stores; 2) Document databases (or stores); 3) Wide-Column (or Column Family) stores; 4) Graph databases [4]. Some popular NoSQL DBMS are listed in the table 01 on basis on their categories.

| Popular NoSQL Databases | | | |
|---|---|---|---|
| **Key-Value stores** | **Document databases/Stores** | **Wide-Column (or ColumnFamily) stores** | **Graph databases** |
| Redis [8] | MongoDB [14] | DynamoDB [11] (Amazon) | Neo4j |
| Riak [9] | ChuchDB [15] | Cassandra [16] (Facebook) | |
| SimpleDB [10] | | Accumulo [17] | |
| Dynamo [11] (Amazon) | | HBase [18] (Apache) | |
| Voldemort [12] (LinkedIn) | | Big Table [19] (Google) | |
| BerkeleyDB [13] Oracle | | Hypertable [19] | |
| | | PNUTS [20] (Yahoo) | |

Table 01: Some popular NoSQL DBMS on basis on their categories.

Through, there are numerous contributions have been made by NoSQL databases; there are still have some bottlenecks for large-scale data management systems. NoSQL DBMSs do not effectively support for applications already written for an earlier generation of RDBMS. To migrate existing applications to adapt to new trends of data growth; to develop new applications on highly scalable OLTP systems, and to rely on existing knowledge of OLTP usage - a new generation of information management systems, termed NewSQL systems. To address big-data OLTP business scenarios that neither traditional OLTP systems nor NoSQL systems address, alternative database systems have evolved, collectively named NewSQL systems.

For Big Data environment, supporting more clients or higher throughput required an upgrade to a larger server. Until recently this meant that implementing a scale-out architecture either required a non-SQL programming model or relying on sharding and explicit replication. There were no solutions that provided complete ACID semantics. This tension is what inspired the NewSQL movement.

## 3 NewSQL

NewSQL is next-Generation Scalable relational database management systems (RDBMS) for Online Transaction Processing (OLTP) that provide scalable performance of NoSQL systems for read-write workloads, as well as maintaining the ACID (Atomicity, Consistency, Isolation, Durability) guarantees of a traditional database system [5],[6],[7],[4]. These systems break through conventional RDBMS performance limits by employing NoSQL-style features such as column-oriented data storage and distributed architectures, or by employing technologies like in-memory processing , symmetric multiprocessing (SMP) or Massively parallel Processing (MPP) advance features and integrate NoSQL or Search components, designed to handle the volume, variety, velocity and variability challenges of Big Data [4].

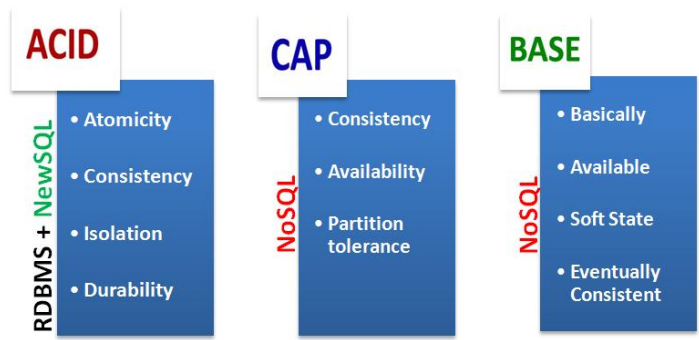

Figure 3.1: ACID, CAP, and BASE properties and supported DMS

Some NewSQL databases are: NuoDB [21] that is a distributed database designed with SQL service: all the properties of ACID transactions, standard SQL language support and relational logic. NuoDB is a web-scale distributed database offering a rich SQL implementation and true ACID transactions. It's also designed from the start as a distributed system that scales the way a cloud service has to scale providing high availability and resiliency. Different from traditional shared-disk or shared-nothing architectures, NuoDB presents a new kind of peer-to-peer, on-demand independence that yields high availability, low-latency and a deployment model that is easy to manage. Designed for the modern datacenter, and as a scale-out cloud database, NuoDB is the NewSQL solution you need to simplify application deployment. ClustrixDB [22] is a distributed SQL database built for large-scale and fast-growing applications. This is a Scale-Out NewSQL Database, for the cloud. ClustrixDB uniquely enables real-time analytics on your live operational data with massively parallel processing. *VoltDB[23]* is an insanely fast in-memory database with incredible high read and write speeds. This NewSQL database supports JSON, event-level transactions. Xeround [24] offers scalable elastic cloud computing infrastructure for elastic data management and data federation within and across clouds. This solution provided from CrunchBase. *MemSQL[25]* is a real-time analytics platform helping companies to query big data quickly and adapt to changing business conditions.

## 4 NewSQL Architecture

This section focuses on the architecture model of NewSQL databases. Traditional databases cannot deliver capacity on demand that application development might be hindered by all the work required to make the database scale. To overcome scalability challenges, developers add scaling techniques like partitioning, sharding and clustering. Another common approach is to add larger machines at more cost. An ideal DBMS should scale elastically, allowing new machines to be introduced to a running database and become effective immediately. Therefore, To adopt scale-out performance, DBMS that has been re-defined relational database technology and implement web-scale distributed database technology to tackle the multiple challenges associated with cloud computing and the rise of global application deployments.

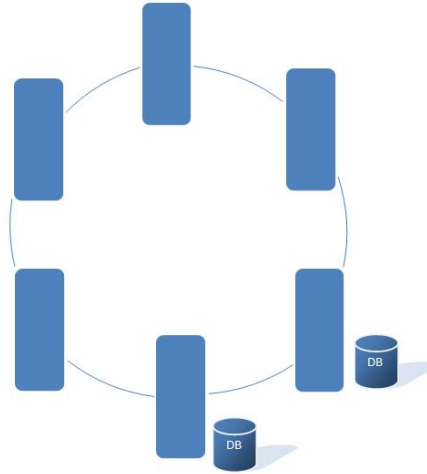

Figure 4.1: web-scale distributed database for web application deployments.

These databases usually distributed architecture in nature and follow three tiers model that split into three layers: an administrative tier, a transactional tier and a storage tier. Traditionally, relational databases were designed for scale-up architectures. In the NewSQL model on-demand scale-out databases stored in distributed datacenters that maintain ACID semantics architecture. Critical key features of these databases are associated with being cloud-scale like ease of provisioning and management, security, agility in high workloads or failures and support for widely distributed applications; in turn, require distributed services that are highly available and can provide low latency. Figure 2.1 shows the architecture of a popular NewSQL database NueDB.

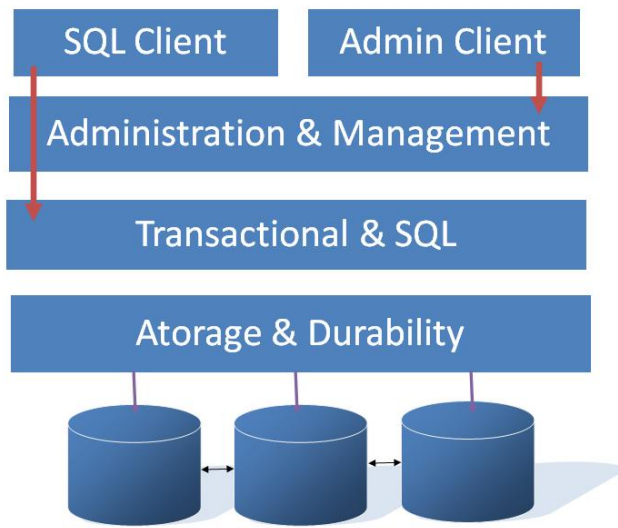

Figure 4.2 shows the architecture of a popular NewSQL database NueDB.

Although NewSQL systems vary greatly in their internal architectures, the two distinguishing features common amongst them is that they all support the relational data model and use SQL as their primary interface [5]. The applications targeted by these NewSQL systems are characterized as having a large number of transactions. These NewSQL systems achieve high performance and scalability by eschewing much of the legacy architecture such as heavyweight recovery or concurrency control algorithms.

## 5 Characteristics of NoSQL Databases

This section describes on the characteristics of NewSQL databases. Technical characteristics of NewSQL solutions are listed in the table 02:

| **Characteristics of NoSQL Databases** | |
|---|---|
| 1 | SQL as the primary mechanism for application interaction. |
| 2 | ACID support for transactions. |
| 3 | A non-locking concurrency control mechanism so real-time reads will not conflict with writes, and thus cause them to stall. |
| 4 | An architecture providing much higher per-node performance than available from traditional RDBMS solutions. |
| 5 | Distributed architecture |
| 6 | A scale-out, shared-nothing architecture, capable of running on a large number of nodes without suffering bottlenecks. |
| 7 | In-memory relational database |
| 8 | Shared-nothing architectures |

Table 02: Characteristics of NoSQL Databases

## 6 Classification of NoSQL Databases

This section describes on the Classification of NewSQL databases. Similar to NoSQL, there are many categories of NewSQL solutions. Categorisation is based on the different approaches adopted by vendors to preserve the SQL interface, and address the scalability and performance concerns of traditional OLTP solutions. NewSQL systems can be categorized into three categories:

| Category of NewSQL Databases | | |
|---|---|---|
| **New Architecture databases** | **New MySQL storage engines** | **Transparent clustering/sharding** |
| VoltDB | TokuDB (commercial) | dbShards (commercial) |
| NuoDB | InfiniDB | ScaleBase (commercial) |
| Drizzle | Xeround | ScalArc |
| Clustrix | GenieDB | Schooner MySQL |
| MemSQL | Akiban | Continuent Tungsten (open source) |
| | | |

Table 03: Classification of NewSQL databases

*New Architecture databases:* These NewSQL systems are newly designed to achieve scalability and performance that are operate in a distributed cluster of shared-nothing nodes. These databases are often written from scratch with a distributed architecture in mind, and include components such as distributed concurrency control, flow control, and distributed query processing. Solutions can be software-only (VoltDB, NuoDB and Drizzle) or supported as an appliance (Clustrix, Translattice). Examples of offerings are Clustrix, NuoDB and Translattice (commercial); and VoltDB, Drizzle, etc., (open source).

*New MySQL storage engines:* The second category are highly optimized storage engines for SQL. MySQL is part of the LAMP stack and is used extensively in OLTP. To overcome MySQL's scalability problems, a set of storage engines are developed. These systems provide the same programming interface as SQL, but scale better than built-in engines, such as InnoDB. Examples of these new storage engines include TokuDB and InfiniDB.

*Transparent clustering/ sharding:* These solutions retain the OLTP databases in their original format, but provide a pluggable feature to cluster transparently, to ensure scalability. Another approach is to provide transparent sharding to improve scalability. Schooner MySQL, Continuent Tungsten and ScalArc follow the former approach, whereas ScaleBase and dbShards follow the latter approach. Both approaches allow reuse of existing skillsets and ecosystem, and avoid the need to rewrite code or perform any data migration. Examples of offerings are ScalArc, Schooner MySQL, dbShards and ScaleBase (commercial); and Continuent Tungsten (open source).

## 7 Comparison of SQL based RDBMS, NoSQL and NewSQL

This section compares SQL based RDBMS, NoSQL and NewSQL databases with set of metrics. Due to performance being the top priority, NoSQL and NewSQL databases tend to have more security gaps than traditional SQL databases.

|  | Traditional RDBMS | NoSQL | NewSQL |
|---|---|---|---|
| **SQL** | Supported | Not supported | Supported |
| **Machine dependency** | Singe machine | Multi-machine/Distributed | Multi-machine/Distributed |
| **DBMS type** | Relational | Non- relational | Relational |
| **Schema** | Table | Key-value, column-store, document store | Both |
| **Storage** | On disk + cache | On disk + cache | On disk + cache |
| **Properties support** | ACID | CAP through BASE | ACID |
| **Horizontal scalability** | Not supported | Supported | Supported |
| **Query Complexity** | Low | High | Very High |
| **Security concern** | Very high | Low | Low |
| **Big volume** | Less performance | Fully supported | Fully supported |
| **OLTP** | Not fully supported | Supported | Fully supported |
| **Cloud support** | Not fully supported | Supported | Fully supported |

Table 04: Comparison of SQL based RDBMS, NoSQL and NewSQL databases

## 8. Conclusion and Future Work

One of the key advances in resolving the "big-data" problem has been the emergence of an alternative database technology. To address big-data OLTP business scenarios that neither traditional OLTP systems nor NoSQL systems address, alternative database systems have evolved, collectively named NewSQL systems. NewSQL is a class of modern relational database management systems (RDBMS) that provide the same scalable performance of NoSQL systems for online transaction processing (OLTP) read-write workloads while still maintaining the ACID guarantees of a traditional database system.

In this paper, we explore background of NoSQL and NewSQL databases systems, discuss some bottlenecks for large-scale data management systems and discuss what inspired the NewSQL movement. This paper covers architecture, characteristics, classification of NewSQL databases for online transaction processing (OLTP) in Big data environment. It also provides the list of popular NoSQL as well as NewSQL databases in separate categorized tables into broad three categories: New Architecture databases; New MySQL storage engines; and Transparent clustering/sharding. Comparison of SQL based RDBMS, NoSQL and NewSQL with metrics.

Naturally, every solution has its pros and cons. Due to performance being the top priority, NoSQL and NewSQL databases tend to have more security gaps than traditional SQL databases. These issues need to research in-depth to overcome the situation. NewSQL database need to be benchmarking, it is important to evaluate "load testing" and "scalability testing" some of popular NoSQL and NewSQL Databases; and compare these databases in respect of Big Data analytics. To compare performance; we need to simulate the exact conditions and workload for processing small and frequent requests and focus on providing fast response times.

# Author

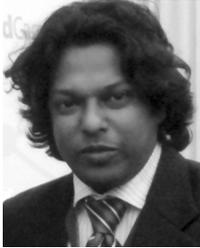

**A B M Moniruzzaman** Received his B.Sc (Hon's) degree in Computing and Information System (CIS) from London Metropolitan University, London, UK and M.Sc degree in Computer Science and Engineering (CSE) from Daffodil International University, Dhaka, Bangladesh in 2005 and 2013, respectively. Currently he is working as a Lecturer of the department of Computer Science and Engineering ad Daffodil International University. He is also working on research on Cloud Computing and Big Data Analytics as a research associate at RCST (Research Center for Science and Technology) at Daffodil International University (DIU), Dhaka, Bangladesh. Besides, his voluntarily works as reviewer of many international journals including IEEE, Elsevier, IGI-Global. He has got 7(seven) international publications including journals and proceedings. He is a student member of IEEE. His research interests include Cloud Computing, Cloud Applications, Open-source Clouds, Cloud Management Platforms, Building Private and Hybrid Cloud with FOSS software, Big Data Management, Agile Software Development, Hadoop, MapReduce, Parallel and Distributed Computing, Clustering, High performance computing, Distributed Databases, NoSQL Databases.